# Anomalous magnetism of Pr in PrCoAsO

Brajesh Tiwari[1], Anand Pal[1,2] and V.P.S. Awana[1]

[1]Quantum Phenomena and Application Division, National Physical Laboratory (CSIR)

Dr. K. S. Krishnan Road, New Delhi-110012, India

[2]Department of Physics, Indian Institute of Science, Bangalore- 560012

**ABSTRACT**

Detailed magnetization and magneto-transport measurements studies are carried out to unearth the anomalous magnetism of Pr in PrCoAsO compound. The studied PrCoAsO sample is single phase and crystallized in the tetragonal structure with space group *P4/nmm* in analogy of ZrCuSiAs type compounds. Detailed magnetization measurements showed that Co moments in PrCoAsO exhibit weakly itinerant ferromagnetic Co spins ordering at below 80 K with a small magnetic moments of ∼ 0.12 $\mu_B$/f.u. High temperatures Curie-Weiss fit, resulted in effective paramagnetic moment $\mu_{eff}$ (exp) of 5.91 $\mu_B$/f.u., which can be theoretically assigned to 3d Co (3.88 $\mu_B$) and 4f Pr (3.58 $\mu_B$). Further, a positive Curie- Weiss temperature (Θ) of 136 K is seen, indicating predominant ferromagnetic interactions in PrCoAsO. Detailed transport measurements showed that PrCoAsO exhibit metallic behavior and negative magneto-resistance below ferro-magnetically (FM) ordered state. Surprisingly, the situation of PrCoAsO is similar to non magnetic La containing LaCoAsO and strikingly different than that as reported for magnetic Nd, Sm and Gd i.e., (Nd/Sm/Gd)CoAsO. The magnetic behavior of PrCoAsO being closed to LaCoAsO and strikingly different to that of (Nd/Sm/Gd)CoAsO is unusual.

# Introduction

The discovery [1] of superconductivity in doped Fe based pnictides REFeAsO (RE = rare earth) ignited the search for the same in other similar structure compounds. The REFeAsO crystallizes in ZrNiCuS type structure, which is an extended family of various 3d metal based pnictides [2]. In this regards the substitution of Fe by Co, in REFeAsO had attracted lot of attention [3-8]. This is primarily because low content (x < 0.20) of Co at Fe site in REFe$_{1-x}$Co$_x$AsO introduces superconductivity [3-5] and secondly the full substitution of Co though not superconducting but results in interesting magnetism [6-10]. For non magnetic La counterpart i.e., LaCoAsO, the Co spins exhibit itinerant ferromagnetism (FM) [6-8], on the other hand for magnetic Nd, Sm and Gd successive PM-FM-AFM transition are reported [8-10]. The successive PM-FM-AFM transitions in case of (Nd,Sm,Gd)CoAsO are explained on the basis of RE[4f] and Co[3d] spins interplay [11]. As far as the Fe counter parts i.e., REFeAsO (RE = La, Sm, Pr, Nd) are concerned, one wonders about the case of Pr, which showed anomalous AFM ordering temperature of around 12 K for Pr spins in PrFeAsO [12]. This is unlike the low temperature usual AFM ordering of RE spins at say 4.5 K in case of SmFeAsO [13] and at 1.8K in NdFeAsO [14]. The unusual AFM ordering of Pr at 12 K in PrFeAsO is explained on the basis of the $Pr^{4f}$ interaction with $Fe^{3d}$ in FeAs plane [12]. This reminds the situation of unusual ordering of Pr at 17 K in cuprate compound $PrBa_2Cu_3O_7$, which has been explained on the basis of $Pr^{4f}$ and $CuO_2$ conduction band hybridization [15].

Though, yet there is some confusion about the origin of 12 K AFM ordering of Pr moments and their exact interaction with adjacent Fe spins, still couple of detailed studies are reported on PrFeAsO and the situation is clear to some extent [12, 16], similar to that as in case of $PrBa_2Cu_3O_7$, where the Pr interacts with Cu-$O_2$ [15]. Keeping in view, the history of unusual ordering of Pr moments in $PrBa_2Cu_3O_7$ [14] and PrFeAsO [12, 16] one wonders how different will be the situation for PrCoAsO. Very recently some of us reported [17, 18] the muon spin spectroscopy and nuclear magnetic resonance (NMR) on RECoPO (RE = La, Pr) and found that Co exhibit weak itinerant FM for both La and Pr. This prompted us to study in detail the magnetic and magneto-transport properties of our PrCoAsO samples and compare the same with reported data on (La,Nd,Sm,Gd)CoAsO [6-10].

PrCoAsO exhibited metallic behavior with Co moments being ordered ferro-magnetically (FM) at below around 80 K, similar to that as for the reported non magnetic LaCoAsO and in sharp contrast to PM-FM-AFM transitions for magnetic (Nd/Sm/Gd)CoAsO. It is surprising Pr being magnetic must resemble more closely with magnetic Nd, Sm and Gd than the non-magnetic La counterpart LaCoAsO. Our results bring out the fact that clearly Pr is anomalous in Fe pnictides series of RECoAsO and the situation could be similar to that as reported earlier for $PrBa_2Cu_3O_7$. Detailed neutron scattering studies are warranted to unearth the unusual magnetism of Pr in various PrFe/CoAsO pnictides.

**Experimental**

Samples of the series PrCoAsO are synthesized through solid state reaction route. High purity (~99.9%) powders of Pr, As, and $Co_3O_4$ in their stoichiometric ratios are properly weighed, mixed and ground thoroughly using mortar and pestle in presence of high purity Ar atmosphere in a glove box (*m-BRAUN*). The mixed powders were palletized in rectangular shape and vacuum-sealed ($10^{-4}$ Torr) in a quartz tube. These sealed quartz tube samples were placed in box furnace heat treated at 550ºC for 12 h, 850ºC for 12 h and then at 1150ºC for 33 h in continuum. Then furnace is allowed to cool naturally. The sintered sample is obtained by breaking the quartz tube. The obtained sample is black in color and slightly brittle. For transport measurements, the obtained sample is again grind and sealed in quartz tube and finally heated at 12 hours at 1150ºC hrs with slow heating rate to obtain a good compact pellet. Finally furnace is cooled slowly down to room temperature. The room temperature x-ray diffraction data is collected by using Rigaku X-ray diffractometer with Cu $K_α$ radiation. The phase purity and lattice parameters are calculated with the help of Rietveld refinement using the FULLPROF SUITE program. The resistivity measurements were performed by a conventional four-point-probe method on a Quantum Design Physical Property Measurement System (*PPMS*-140kOe). Magnetic and Magneto-transport properties of the samples were carried out with same *QD-PPMS* in the temperature range 2-300 K and in field up to 10 Tesla.

**Results and Discussion**

Figure 1 shows the Rietveld fitted x-ray diffraction (XRD) pattern of the studied PrCoAsO sample. It is clear from Figure 1 that the samples is single phase and crystallized in the tetragonal structure with space group *P4/nmm* in analogy of ZrCuSiAs, except a few minor peaks close to the background. *FULLPROF SUITE* program has been used to refine the room temperature X-ray diffraction pattern of the studied samples. The Rietveld analysis is carried out in the space group P4/*nmm* (#129) with Wyckoff positions for Pr and As taken to be located at 2*c* (1/4, 1/4, *z*), the Co at site 2*b* (3/4, 1/4, 1/2) and the O at 2*a* (3/4, 1/4, 0). The refined lattice parameters for PrCoAsO are a = 4.013(1) Å, c = 8.354(2) Å with fractional atomic positions for Pr and As at (1/4, 1/4, 0.1535) and (1/4, 1/4, 0.627) respectively. The representative unit cell of PrCoAsO is shown in inset of Figure 1.

The magnetization i.e. magnetic moment versus temperature plot for PrCoAsO sample is depicted in Figure 2. It is clear that the Co moments exhibit weakly itinerant/localized ferromagnetic Co spins at below 80 K with a small magnetic moments of M = 0.12 $\mu_B$/f.u. Upper inset of Figure 2 shows the isothermal magnetization [M (H)] curves at temperatures 100 K, 50 and 10 K. Besides the FM contribution from Co spins, the linear component from Pr paramagnetic moments is also seen, as magnetization increases linearly at high fields (up to 10 kOe). This is unlike LaCoAsO, for which M(H) is comprised of purely the Co FM moments and hence near complete saturation of moments is reported below FM ordering temperature [6, 19, 20]. It is clear from Figure 2 that Pr possess finite paramagnetic moment in PrCoAsO, the exact value of the same could be detected from the high temperature magnetic susceptibility data. From high temperatures Curie-Weiss $\chi = C^{mol}/T - \Theta$ fit, we found effective paramagnetic moment $\mu_{eff}$ (exp) = 5.91 $\mu_B$/f.u., which is close to theoretical, if one assumes that contribution to magnetic moment is from 3d Co (3.88 $\mu_B$) and 4f Pr (3.58 $\mu_B$) i.e. $\mu_{eff}^{PrCoAsO} = \sqrt{\mu_{Co}^2 + \mu_{Pr}^2} \mu_B = 5.28\ \mu_B$ with Curie constant $C^{mol}$ = 4.38 emu. mol$^{-1}$.Oe$^{-1}$K. Further positive Curie- Weiss temperature ($\Theta$ = 136 K) is seen, indicating predominant ferromagnetic interactions. It is observed that I$\Theta$I /T$_c$ (~1.7) is away from the value unity, which can be attributed to the presence of different type of exchange interactions in PrCoAsO. Magnetization below T$_C$ (50 K and 10 K) should show saturated S shape curves for ferromagnetic ordering as expected from magnetization versus temperature curve and positive Curie- Weiss constant. However, there is no sign of magnetic saturation, indicating presence of antiferromagnetic coupling between magnetic moments weakly itinerant 3d and localized 4f electrons as reported earlier [21, 22]. One can see from Figure 2 a small change in slope at around 250 K from 1/$\chi$ vs T curve possibly due to some short range ordering, which ultimately develops a long range ferromagnetic order at 80 K. First derivative of magnetic susceptibility shows a clear transition at 62 K (lower inset of Figure 2), indicating that bulk FM order is settled below this temperature. It is clear from magnetization results of Figure 2, that Co spins undergo mostly single FM order below say 80 K and Pr possess definite paramagnetic moment in PrCoAsO. However, for Nd/Sm/GdCoAsO the successive PM-FM-AFM transitions are reported widely [8-11, 21]. As mentioned in introduction itself the Co itinerant FM in case of LaCoAsO [6, 19, 20] and successive PM-FM-AFM transition for Nd/Sm/GdCoAsO [8-11, 21] are already reported and explained in terms of RE[4f] and Co[3d] interplay [8-11]. The absence of successive PM-FM-AFM transition for magnetic RE (Pr) containing PrCoAsO is surprising and warrants clear cut explanations.

The resistivity versus temperature plots of the PrCoAsO sample in applied fields of up to 100 kOe show metallic behavior with a clear anomaly upon magnetic ordering as can be seen in Figure 3 in temperature range of 2- 300 K. A linear dependence of resistivity can be seen for high temperatures (200 -300 K) which suggests resistivity is dominated by electron-phonon interaction with a clear anomaly around 60 K which is characteristic of ferromagnetic ordering. To see the effect of applied magnetic fields on anomaly in resistivity, first derivative of resistivity (dρ/dT) as a function of temperature (T) are shown as inset of Figure 3 for clarity. A clear increase in the temperature of resistive anomaly can be observed from 60 K in absence of magnetic field to 89 K in the external field of 100 kOe, in conjugation with smearing (broadening) of transition. Strong temperature dependence and smearing of this anomaly upon external magnetic field suggests a complex interplay of charge, spin and lattice degrees of freedom in PrCoAsO system. Broadening of Co-ordering transition can be attributed to polarization of Pr magnetic moments upon external magnetic field which can be manifestation of increased 3d-4f interaction. Interestingly, PrCoAsO does not show field driven resistivity upturns like in case of Sm/Nd/GdCoAsO [8-11, 21]. Field driven resistivity upturns in case of Sm/Nd/GdCoAsO are seen due to successive field dependent PM-FM-AFM transitions [8-11, 21]. Field driven resistivity up turns are obviously missing in case of PrCoAsO, which undergoes a single FM transition and hence its magneto-transport data are close to LaCoAsO [6, 19, 20] than the Sm/Nd/GdCoAsO [8-11, 21].

Isothermal magneto-resistance defined as $\% MR = \rho(H) - \rho(0)/\rho(0)$ where ρ(H) and ρ(0) are the resistivity in presence and absence of field measured up to applied magnetic field of 100 kOe as depicted in Figure 4. Just below Co ordering temperature a maximum of -6 % MR is observed at 50 K, while it reduces away from transition in paramagnetic and ferromagnetic phases similar to LaCoAsO [6, 19, 20] but unlike other RECoAsO [8-11, 21]. It is important to notice that low field (< 10 kOe) magneto-resistance is linear in ferromagnetic phase in contrast to paramagnetic phase. Due to delocalization of electrons in presence of low external magnetic fields, conductivity increases quadraticaly as this is the case of paramagnetic phase. Upon ferromagnetic ordering a linear magneto-resistance at low fields may be due to small effective mass of electrons at Fermi surface which need to be probed on single crystals of PrCoAsO as it is not possible to determine on polycrystalline sample which may also lead to linear magneto-resistance in polycrystalline metals at high fields due to inhomogeneity. The isothermal magneto-resistance (MR) behavior of PrCoAsO is similar to that as reported for LaCoAsO [6, 19, 20] but much different than as for Sm/Nd/GdCoAsO [8-11, 21]. It is interesting that in RECoAsO series, the magnetic Pr follows more closely the non-magnetic La than the magnetic Sm and Nd counterparts. The occurrence of field driven resistivity upturn in Sm/Nd/GdCoAsO is discussed in detail earlier [8-11, 21] and the origin of same is attached again to complex interaction of Nd moments with the Co in CoAs/P layer. In case of Fe counter part of PrCoAsO, i.e., PrFeAsO, a 12 K resistivity step is reported earlier in resistivity data and was assigned to the unusual ordering of Pr moments due to $Pr^{4f}$ and $Fe^{3d}$ interaction [12, 16]. The issue here is that PrCoAsO behaves more closely like LaCoAsO than the magnetic Sm/NdCoAsO. This is intriguing because unlike non magnetic La, Pr must behave more closely like the magnetic Nd and Sm counterparts. In fact the Co FM ordering is same for both LaCoAsO and PrCoAsO, as if $Pr^{4f}$ moments do not affect the Co spins at all.

In conclusion, this study brings out the fact that Pr exhibits anomalous behavior in PrCoAsO, as the same though magnetic but still behaves more closely with non magnetic La counterpart LaCoAsO than the magnetic RE comprising Sm/Nd/GdCoAsO. The complex magnetic structure of studied PrCoAs compound needs to be unearthed. One needs to carry out detailed neutron scattering experiments to know exactly the Pr and Co spin arrangements in case of PrCoAsO.

**Acknowledgement:**

Authors would like to thank their Director Prof. R.C. Budhani for his keen interest in the present work. This work is supported by DAE-SRC outstanding investigator award scheme to work on search for new superconductors.

# Figure Captions

Figure 1: Room temperature observed and fitted x-ray diffraction pattern for PrCoAsO along with Bragg positions. Inset: represents the unit cell of PrOAsO with their atom type indicated there in.

Figure 2: Magnetization (M) versus temperature (T) for PrCoAsO at 50 Oe. $1/\chi$ vs T graph is shown in right y-axis with Curie-Weiss as solid line in red. Insets show the M(H) curve at different temperatures (10 K, 50 K and 100 K) and derivative of magnetic susceptibility with clear $T_C$ at 62 K.

Figure 3: The resistivity ($\rho$) versus temperature (T) for PrCoAsO at different fields respectively from 300 down to 2.5 K. Inset: first derivative of resistivity (d$\rho$/dT) as function of temperature show clear shift in anomaly in resistivity with applied magnetic field as indicated by a solid arrow.

Figure 4: Magneto-resistance (MR) at different temperatures (5 - 200 K) upto applied magnetic field of 100 kOe.

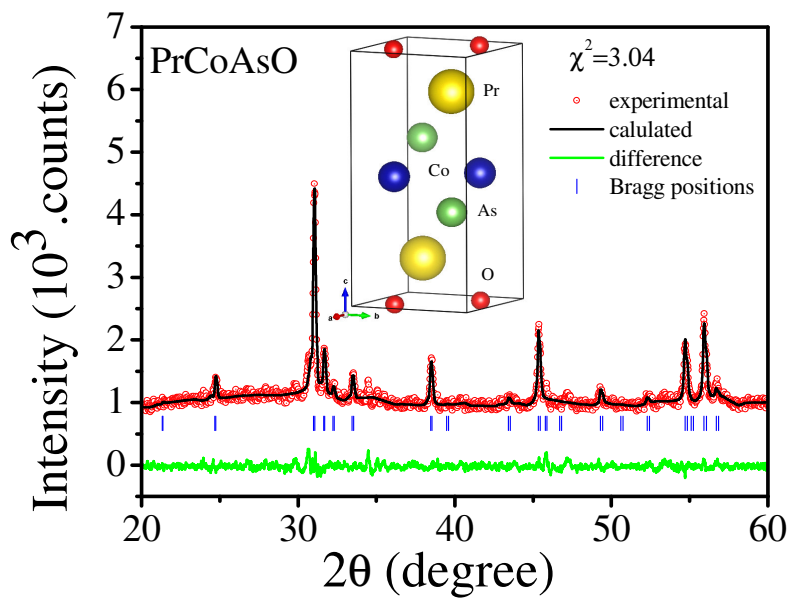

Figure 1

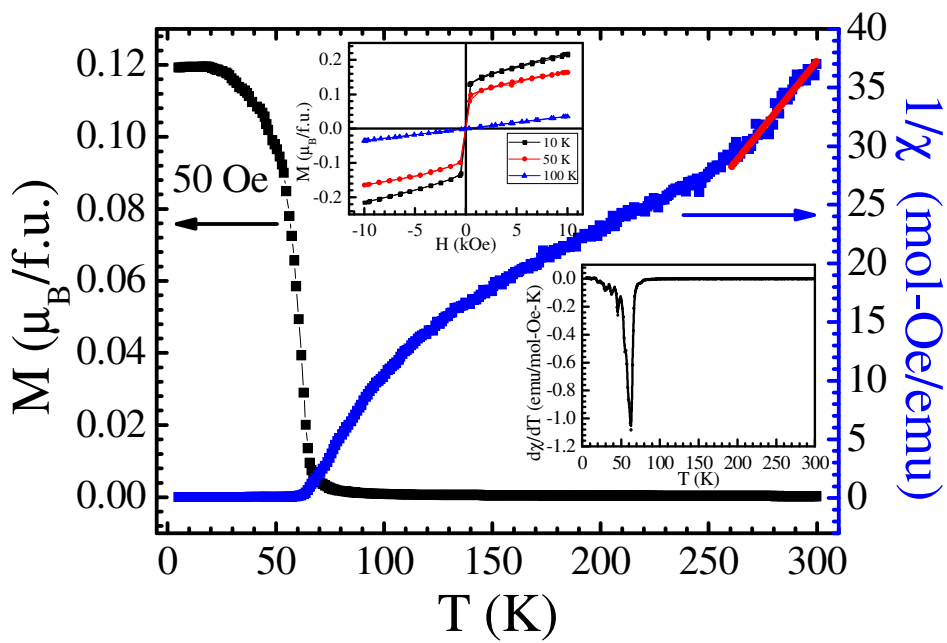

Figure 2

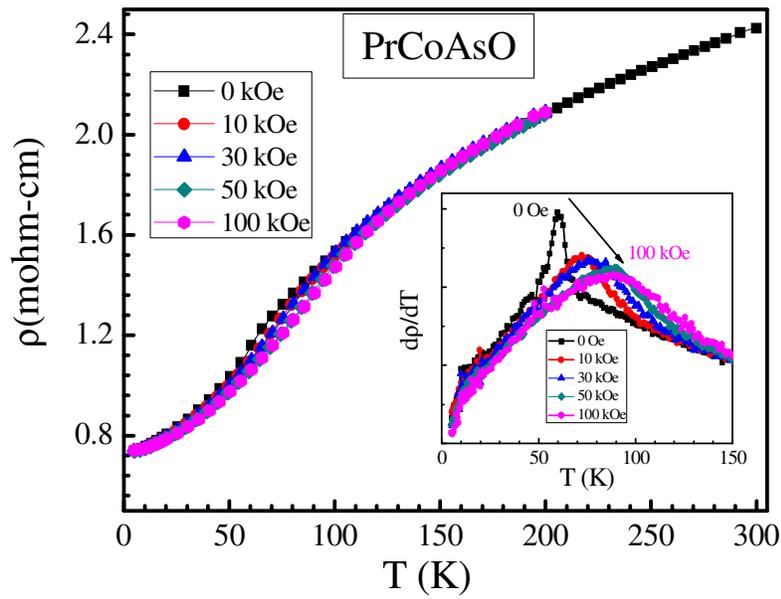

Figure 3

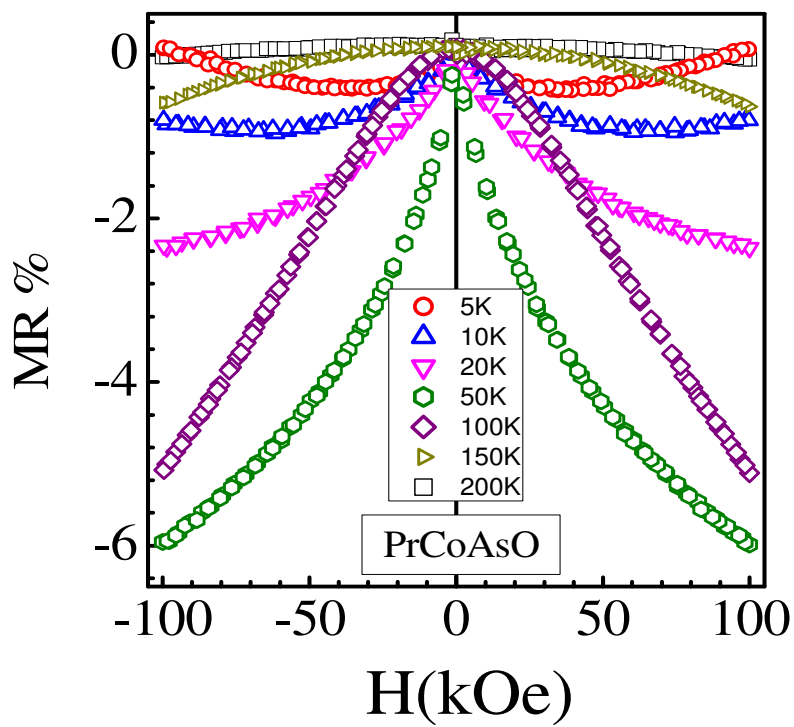

Figure 4

8